\documentclass{article}

\usepackage[english]{babel}

\usepackage[letterpaper,top=2cm,bottom=2cm,left=3cm,right=3cm,marginparwidth=1.75cm]{geometry}

\usepackage{amsmath}
\usepackage{graphicx}
\usepackage{algorithm}
\usepackage[noend]{algorithmic}
\algsetup{linenosize=\small}
\usepackage[colorlinks=true, allcolors=blue]{hyperref}

\title{Towards the Maximum Traffic Demand and Throughput Supported by Relay-Assisted mmWave Backhaul Networks}
\author{Qiang Hu, Yuchen Liu, Yan Yan, Miao Liu, Jun Zheng, and Douglas M. Blough}

\begin{document}
\maketitle

\begin{abstract}
This paper investigates the throughput performance issue of the relay-assisted mmWave backhaul network. The maximum traffic demand of small-cell base stations (BSs) and the maximum throughout at the macro-cell BS have been found in a tree-style backhaul network through linear programming under different network settings, which concern both the number of radio chains available on BSs and the interference relationship between logical links in the backhaul network. A novel interference model for the relay-assisted mmWave backhaul network in the dense urban environment is proposed, which demonstrates the limited interference footprint of mmWave directional communications. Moreover, a scheduling algorithm is developed to find the optimal scheduling for tree-style mmWave backhaul networks. Extensive numerical analysis and simulations are conducted to show and validate the network throughput performance and the scheduling algorithm.
\end{abstract}

\section{Introduction}

Facing the shortage of available spectrum in microwave bands and the increasing 
global mobile data traffic demand, millimeter-wave (mmWave) technology obtains 
a lot attention and is being heavily researched due to its potential of enabling 
broadband radio access and backhaul with ultra-high data speed in future 5G 
wireless networks~\cite{pi2011introduction, weiler2014enabling, 
	verma2015backhaul, jaber20165g, yan2021load}.
	
In the 5G and B5G (beyond 5G) era, the cell size becomes smaller, and operators will deploy a 
large number of small-cell base stations (BSs), which cooperate with legacy 
macro-cell BSs, to provide better coverage and higher data rate to cellular 
users in urban environments. However, most of small-cell BSs may not have 
wired connections to the core network due to the construction limits and 
prohibitive cost in many places such as North America, so that small-cell 
BSs can only connect to the core network with the help from nearby macro-cell 
BSs acting as gateway nodes. Thus, this small-cell trend brings the 
\textit{backhaul challenge} of transferring the huge amount of data between 
small-cell BSs and macro-cell BSs. Being able to support multi-Gbps wireless 
links, mmWave backhauling has been regarded as a promising solution to the 
backhaul challenge in 5G cellular systems.

To make the mmWave backhaul network feasible, all these unique propagation 
features of mmWave signals, including the high path loss~\cite{frey1999effects} 
and the blockage effect~\cite{liu2019analysis, singh2009blockage}, and the development of mmWave
communication technologies have to be taken into account. Therefore, We proposed 
a relay assisted mmWave network architecture for backhaul in urban areas~\cite{
hu2017relay,hu2018optimizing, hu2020feasibility, yan2018path, yan2021feasibility, liu2020joint}, 
in which mmWave relay 
nodes are used to assist in connecting BSs of the wireless network. Although 
the idea of relaying appeared in the literature for a long time in both legacy 
cellular systems~\cite{yang2009relay,peters2009relay} and the general mmWave networks~\cite{lan2011space,zheng2016toward,niu2015blockage,qiao2012efficient},
we are the first to introduce dedicated mmWave relays to help build mmWave
backhaul networks with the ultra-high throughput requirement.

In our previous work~\cite{hu2020feasibility}, we have explored the methods to select 
mmWave relays from a set of candidate relay locations, so that 
interference-minimal logical links between BSs can be constructed to 
support the targeting high volume backhaul traffic in the urban area. 
However, our previous work mainly focuses on addressing the feasibility 
issue of constructing the relay-assisted mmWave backhaul network, 
while few detail has been provided from the perspective of performance. 
After the relay-assisted mmWave backhaul network has been built 
(i.e., all logical links have been constructed), to find out the maximum 
traffic demand of small-cell BSs that could be supported by the 
backhaul network is of interest to us. 

Finding an efficient scheduling scheme is not trivial under different 
mmWave backhaul network settings. Specifically, there are two critical 
constraints that affect the analysis on the throughput performance of a 
mmWave backhaul network. 
The first one is the number of radio chains available on each BS. 
It is assumed that one radio chain cannot transmit and 
receive at the same time, which is referred to as the primary interference. 
One radio chain can only serve the transmission on one logical link at a time.
If there are not ``enough" radio chains available, the BS may not
be able to simultaneously transmit data on all its attached logical links. 
In fact, in a multi-hop mmWave backhaul network, most of the BSs have more 
than one logical links attached; thus, 
it is meaningful to investigate the impact of limited radio chain resources 
on the throughput performance of the backhaul network. 
The other constraint is rooted in the possible existence of secondary mutual 
interference among different logical links in the backhaul network. 
In~\cite{hu2020feasibility}, we try to construct a secondary mutual 
interference-minimal network topology; however, the simulation results show 
that it may not be always possible to form such a backhaul network with the 
interference-minimal property through relay selection. 
In the case where mutual interference exists, to maintain the high throughput 
of logical links, we have to schedule the transmissions of those interfering 
logical links into different time periods, so that they do not affect each
other. Therefore, it is interesting to investigate the throughput performance 
of a relay-assisted mmWave backhaul network where limited secondary 
mutual interference exists between a few pairs of logical links. 

In this paper, we formulate several optimization problems to explore the 
throughput performance of a relay-assisted mmWave backhaul network 
considering two different traffic models. First, we assume each small-cell 
BS maintains the same traffic demand, and linear programming is used to
maximize that traffic demand. Second, if small-cell BSs have 
different traffic demand, instead, the aggregated traffic demand at the 
macro-cell BS in the network is maximized. 
Each problem is categorized into four individual cases according to 
the two factors discussed above, i.e., whether there are enough radio chain 
resources on BSs and whether limited secondary 
mutual interference exists between logical links. Moreover, we propose a 
simple scheduling algorithm that can be used to implement a schedule where 
each logical link can achieve its schedule length obtained through 
solving the optimization problems. Simulation results are provided 
to show the throughput performance of our proposed relay-assisted 
mmWave backhaul networks in different cases.

\section{Related work}
mmWave backhaul networks and their performance evaluation have attracted a significant amount of interest in the research academia recently. In~\cite{arribas2019optimizing}, the authors study the mmWave self-backhaul scheduling problem and derive an MILP formulation for it as well as upper and lower bounds. They prove that the problem is NP-hard and can be approximated, but only if interference is negligible. 
Given a set of mm-wave backhaul links, ~\cite{saad2019millimeter} addresses the problem of backhaul scheduling through considering the mutual interference and number of radio chains constraints. A succinct optimization-based formulation of the problem is provided and using reduction from the set-cover problem, the authors devise a provably good polynomial-time algorithm for the problem.
The authors of~\cite{niu2019relay} investigate the problem of optimal scheduling to maximize the number of flows satisfying their QoS requirements with relays exploited to overcome blockage, where relays refer to small-cell BSs. Both a relay selection algorithm and a transmission scheduling algorithm are proposed to increase the network throughput through addressing the blockage issue and exploiting concurrent transmissions. 
Note that all the works above consider the self-backhaul network where mmWave small-cell BSs serve as relays, and none of them take dedicated mmWave relays as network elements into account. In contrary, when dedicated relays are introduced in our network model, the scheduling problem becomes more sophisticated. Moreover, we also propose a new limited-interference model which can not only captures the mutual interference feature in urban mmWave backhaul networks, but also be used to simplify the formulation of such complex scheduling problem.

The authors in~\cite{fang2021joint} optimize the scheduling of access and backhaul links such that the minimum throughput of the access links is maximized based on the revised simplex method. ~\cite{ranjantwo} considers the radio resource scheduling problem and presents a QoS-based downlink scheduler designed explicitly for integrated access and backhaul (IAB) networks. The scheduler is devised after considering multihop relaying topology, QoS requirements and backhaul constraints. These works are different from ours because they consider an IAB network structure whereas the frequency band used for backhaul in our research is dedicated to the backhaul usage.

In~\cite{yuan2018optimal}, the authors propose an efficient scheduling method, so-called schedule-oriented optimization, based on matching theory that optimizes QoS metrics jointly with routing in the mmWave cellular network. It is claimed to be capable of solving any scheduling problem that can be formulated as a linear program whose variables are link times and QoS metrics. As an example, they also show the optimal solution of the maximum throughput fair scheduling (MTFS). However, this work does not consider a relay assisted backhaul network architecture, and focuses on the joint optimization of both QoS metrics and routing.

More recently, due to the fast development of artificial intelligence, several learning based scheduling methods for mmWave networks are also proposed. To achieve resilience to link and node failures, the authors in~\cite{dogan2021reinforcement} explore a state-of-the-art Soft Actor-Critic (SAC) deep reinforcement learning algorithm, that adapts the information flow through the network, without using knowledge of the link capacities or network topology. This work uses an N-relay Gaussian Full-Duplex (FD) 1-2-1 network model which is different from our network model. ~\cite{zhang2021resource} proposes a reinforcement learning approach based on column generation to address the resource allocation problem for mmWave IAB backhaul, as the method is claimed to be able to capture the environment dynamics such as moving obstacles in the network.

\section{System model}
\begin{figure}[ht]
  \centering
  \includegraphics[width=0.9\linewidth]{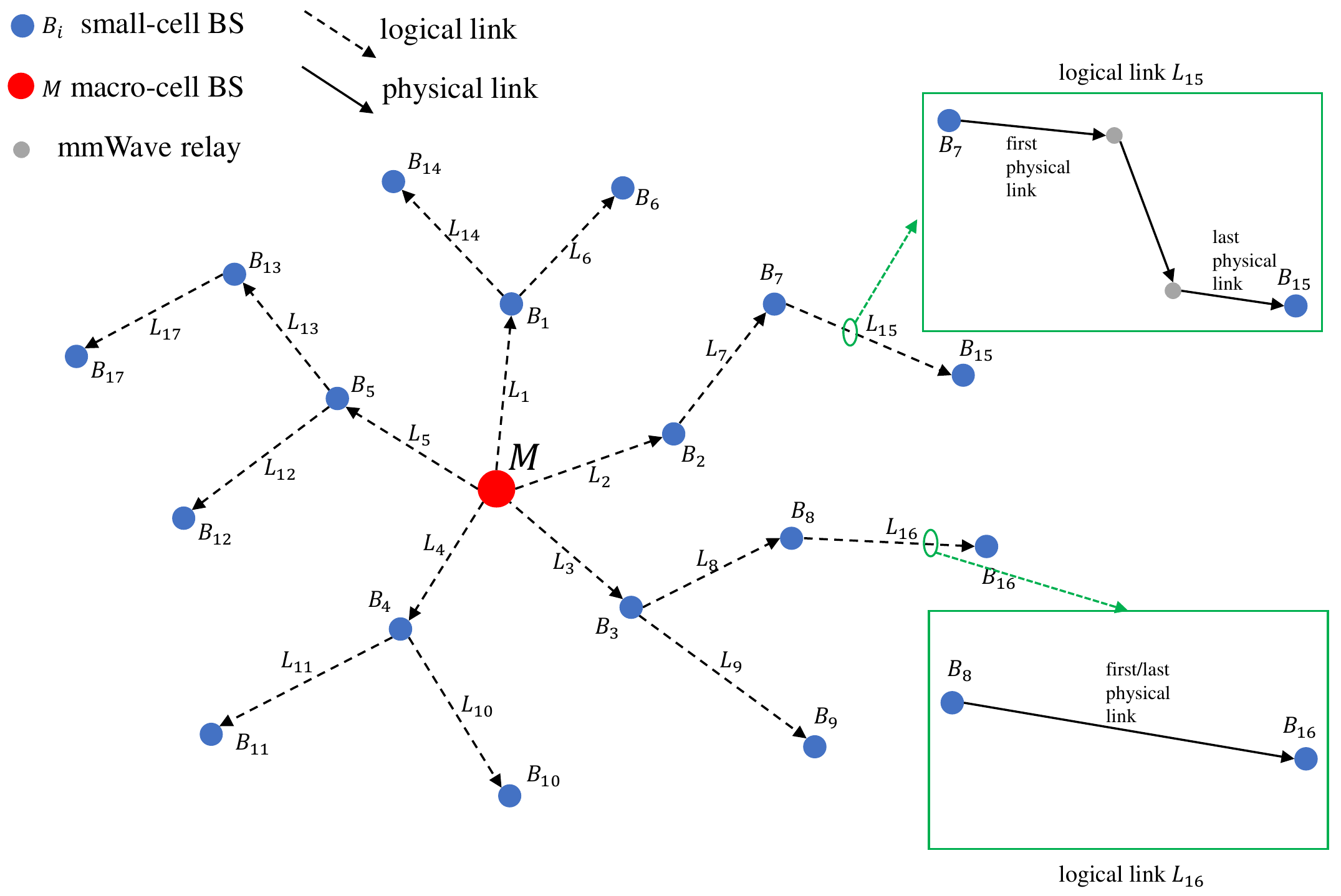}
  \caption{An example of a relay-assisted mmWave backhaul network with tree topology (arrows indicate the traffic direction of links in the downlink case)}
  \label{fig:topology_backhaul_ch5}
\end{figure}

We consider that a relay-assisted mmWave backhaul network has already 
been constructed and given to us. The backhaul network contains a single 
macro-cell BS $M$, a set of small-cell BSs 
$\mathcal{B}=\{B_1, B_2, ...\}$, and a set of mmWave relays. 
The backhaul network has a tree topology that defines a set of backhaul 
logical links $\mathcal{L}=\{L_1, L_2, ...\}$ which connect all BSs. 
In the tree topology, we define the single ``inbound" logical link between 
$B_i$'s parent BS and BS $B_i$ as $L_i$. In Figure~\ref{fig:topology_backhaul_ch5}, an example
of a relay-assisted mmWave backhaul network with a tree topology containing
17 logical links is shown.

Note that since the downlink (i.e., from macro-cell 
BS to small-cell BSs) and uplink (i.e., from small-cell BSs to macro-cell 
BS) traffic in the backhaul network can be easily duplex-ed in 
the time domain or frequency domain (e.g., the Time Division Duplex (TDD) and 
the Frequency Division Duplex (FDD) used in 4G-LTE systems~\cite{yonis2012lte,ku2011resource,scheme2009lte}), 
we optimize the traffic demand of small-cell BSs in either the downlink or 
the uplink case for simplicity.
Note that, to obtain the optimal traffic demand of small-cell BSs in the 
downlink and uplink hybrid case is closely related to the general scheduling 
problem in the wireless network, which is usually considered as NP-complete 
depending on the specific mutual interference model in use, and we leave it as 
future work.
The optimization process described in this chapter uses the downlink case
as an example, despite that the same optimization formulation can be applied
to the uplink case as well. Therefore, all logical links and physical links 
defined in the backhaul network are directional. A logical link is either 
a single-hop LoS physical link or a multi-hop path going through several 
relays sequentially from the source BS to the destination BS. 
In the downlink backhaul scenario, as the macro-cell BS serves as the 
gateway node to the backbone Internet for all small-cell BSs, all the 
data traffic requested from each small-cell BSs comes from the macro-cell 
BS, and the mmWave backhual network disseminates data traffic 
from the macro-cell BS to every small-cell BS.  

To maintain the high throughput capability of a logical link, we assume 
that all physical links within a logical link are mutual 
interference-minimal, which is referred to as the ``intra-path" 
interference-minimal. Note that, ``interference-minimal" here means that 
the amount of mutual interference has been controlled to a minimal level 
which is even smaller than the noise level, so that in the latter analysis, 
that minimal interference can be neglected. For each intra-path 
interference-minimal logical link, we can obtain its optimal schedule for 
all its physical links according to Theorem 1 in our previous work~\cite{hu2017relay}. 
Therefore, we can further get the corresponding maximum end-to-end capacity 
$C_i$ of a logical link $L_i\in \mathcal{L}$. To achieve $C_i$ of $L_i$, 
the first physical link of $L_i$ is scheduled to transmit for a portion
$P_i^f\in (0,1]$ of the total schedule length, while the portion of the 
scheduled transmission for its last physical link in the total schedule 
length is denoted as $P_i^l\in (0,1]$. 

As mentioned before, the number of radio chains $N_i^R$ on a small-cell BS $B_i$ 
($N_M^R$ on the macro-cell BS $M$) is a key factor that affects the scheduling 
of all logical links attached to the same BS. According to the definition 
in~\cite{coleman2018cwna}, ``a radio chain is defined as a single radio and 
all of its supporting architecture, including mixers, amplifiers, and 
analog/digital converters." We assume that radio chains considered in this work 
are half-duplex, which means a radio chain will not transmit and receive at the 
same time. As for the multiple logical links attached to a BS with single radio 
chain, they cannot be active, either transmitting or receiving, simultaneously. 
This is usually referred to as the primary interference. 
However, when multiple radio chains are available on a BS, if the mutual 
interference between logical links does not exist (i.e., interference-minimal), 
some of them could be active at the same time. 
Note that, since mmWave relays are devices much simpler and cheaper than BSs, 
each relay only has one radio chain; thus relays are always constrained by the 
primary interference.

\subsection{New model of mutual interference between logical links}
TDMA-based scheduling in wireless networks has been researched a lot in the literature. 
If any two backhaul logical links are possible to interfere with each other, depending 
on the exact interference model in use, the general scheduling problem is often 
NP-complete~\cite{ramanathan1997unified,ergen2010tdma,yi2008complexity}. 
Sometimes it is even hard to find polynomial approximate algorithm to address the scheduling problem~\cite{sharma2006complexity,moscibroda2007optimal,blough2008framework,blough2010approximation}. 

However, due to the specific propagation features of mmWave signals, such as the well-known 
blockage effect and directional transmissions, the interference relationship is not that 
``general" in a relay-assisted mmWave backhaul network in the dense urban environment. 
Based on the simulation results in~\cite{hu2020feasibility}, in the cases where the 
interference-minimal backhaul network is not feasible through using our proposed relay 
selection algorithm, we find that the path searching for those problematic logical links 
usually ``fails" due to the searched path of a problematic logical link is interfering with 
some already constructed logical links which intersect with the problematic logical link.

Based on the above observation, for the relay-assisted mmWave backhaul network, 
we propose a new mutual interference model that only logical links connecting to 
the same BS would interfere with each other and a logical link can at most 
interfere with one other logical link at a BS. In the worst case, a logical link 
may interfere with two logical links in total, one at each end. This model is 
reasonable because the ultra high throughput requirement of a mmWave backhaul 
network requires the mutual interference to be controlled to a minimal or close-to 
minimal level, and if too many logical links interfere with each other, the 
throughput performance will be very bad. Moreover, we also observe that after 
modifying the relay selection algorithm in~\cite{hu2020feasibility} which 
allows limited mutual interference exists between logical links as described in 
the new interference model, the feasibility of finding high throughput 
relay-assisted mmWave backhaul network increases a lot. The further optimization 
problem formulations in this chapter are using the new mutual interference model.


\section{Maximizing the backhaul throughput performance}
\label{sec:ch6:max_traffic}
In this section, we are going to address the issue of finding both the maximum traffic 
demand of small-cell BSs and the maximum backhaul throughput at the macro-cell 
BS of a given relay-assisted mmWave backhaul network. To answer these two questions, 
we have to find out the corresponding optimal schedule of each logical link in the 
network with two different traffic models, respectively. Therefore, we formulate the 
scheduling problem using linear programming, which can be solved efficiently, under 
different backhaul network configurations.

The traffic demand of a small-cell BS $B_i$ is defined as $D_i$ and the overall
backhaul throughput at the macro-cell BS $M$ is denoted as $D_M$. Thus, the 
relationship between $D_i$ and $D_M$ is
\begin{equation}
    D_M = \sum_{i=1}^{N_B}{D_i}
\end{equation}
where $N_B$ is total number of small-cell BSs in the backhaul network.

\subsection{Maximizing the traffic demand of small-cell BSs}
\label{sec:max_td_sbs}
We first address the problem of maximizing the traffic demand of small-cell BSs
in the backhaul network. In this problem, we assume each small-cell BS maintains
the same traffic demand $D_B$, i.e., $D_i = D_B,\ \forall\ i \in \{1,2, ..., N_B\}$. 
In practice, the traffic demand of a small-cell BS is determined by the actual 
amount of data requested from all user ends that access to it, which is quite relevant to 
the location, time, and other environmental factors. Despite that the traffic demand 
may be different among different small-cell BSs, it is still reasonable to make 
the above assumption, as we aim to maximizing the traffic demand of each small-cell BS
without taking the specific environmental factors into consideration. 

We categorize the optimization problem into four different cases according to the different 
conditions on the existence of mutual interference between logical links and the number of 
available radio chains on BSs. Each case will be discussed in detail later. 

\subsubsection{Case 1: interference-minimal, enough-radio-chain}
In this case, the backhaul 
network is interference-minimal, which means that any two logical links scheduled to 
be active concurrently will not experience end-to-end throughput decrease due to the 
factor of mutual interference. Moreover, the meaning of ``enough radio chains" is that 
the numbers of radio chains $N_i^R$ on a small-cell BS $B_i$ as well as $N_M^R$ on the 
macro-cell BS $M$ are no fewer than the number of logical links attached to $B_i$ and 
$M$, respectively. In this scenario, different logical links can be scheduled to 
transmit data simultaneously without being constrained by the primary interference, 
because each logical link has its own radio chain.

Therefore, we only need to make sure that the single ``inbound" logical link $L_i$ at 
the BS $B_i$ has enough link capacity to support the total traffic demand of all 
small-cell BSs in the sub-tree rooted at $B_i$. This is called the 
\textit{logical link capacity constraint}. The optimization problem of this case can 
be written as,
\begin{equation}
\begin{split}
    \max \quad & D_B \\ 
    \mathrm{s.t.} \quad & C_i \geq \sum_{B_j \in \mathcal{B}_{i}}{D_j} = |\mathcal{B}_{i}| D_B,\ \forall\ i \in \{1,2, ..., N_B\} \\
\end{split}
\end{equation}
where $\mathcal{B}_{i}$ is the set of small-cell BSs in the sub-tree rooted 
at $B_i$, and $|\cdot|$ is the operation which returns the number of elements in a set.

\subsubsection{Case 2: interference-minimal, limited-radio-chain}
\label{sec:lp_case2}

In this case, we move a step forward, and consider that there are not ``enough" radio 
chains available on BSs. It means that multiple logical links connecting to a BS have 
to compete for the use of available radio chains, because one radio chain can only 
serve one active logical link at a time. 
Note that one logical link cannot use multiple radio chains to transmit or receive data 
simultaneously due to the interference issue. Meanwhile, the mutual interference is still 
controlled to the minimal level in the backhaul network, which is the same as that in the 
case 1. 

Therefore, it is found that the total active time of all physical links attached to a 
small-cell (macro-cell) BS $B_i$ ($M$) should not exceed $N_i^R$ ($N_M^R$) times the 
total schedule length, because every radio chain on a BS could be actively working for 
the entire schedule length if the transmissions on the logical links are properly 
scheduled in the ideal case. This is referred to as the \textit{limited radio chain 
constraint}. 

As for a physical link attached to a BS, it is either the first physical link of a 
logical link starting at the BS or the last physical link of a logical link ending 
at the BS. Since every physical links within a logical link follows the optimal 
schedule of that logical link, the scheduled active time for the first and last 
physical link of the logical link $L_i$ is no longer than $P_i^f$ and $P_i^l$ of 
the total schedule length, respectively. It is because that $P_i^f$ or $P_i^l$ is 
the maximum schedule length for the first or last physical link in $L_i$, which can
only be achieved when the logical link $L_i$ is scheduled to be active during the 
entire schedule length. We called it the \textit{physical link scheduling constraint}.

\begin{figure}[ht]
  \centering
  \includegraphics[width=0.9\linewidth]{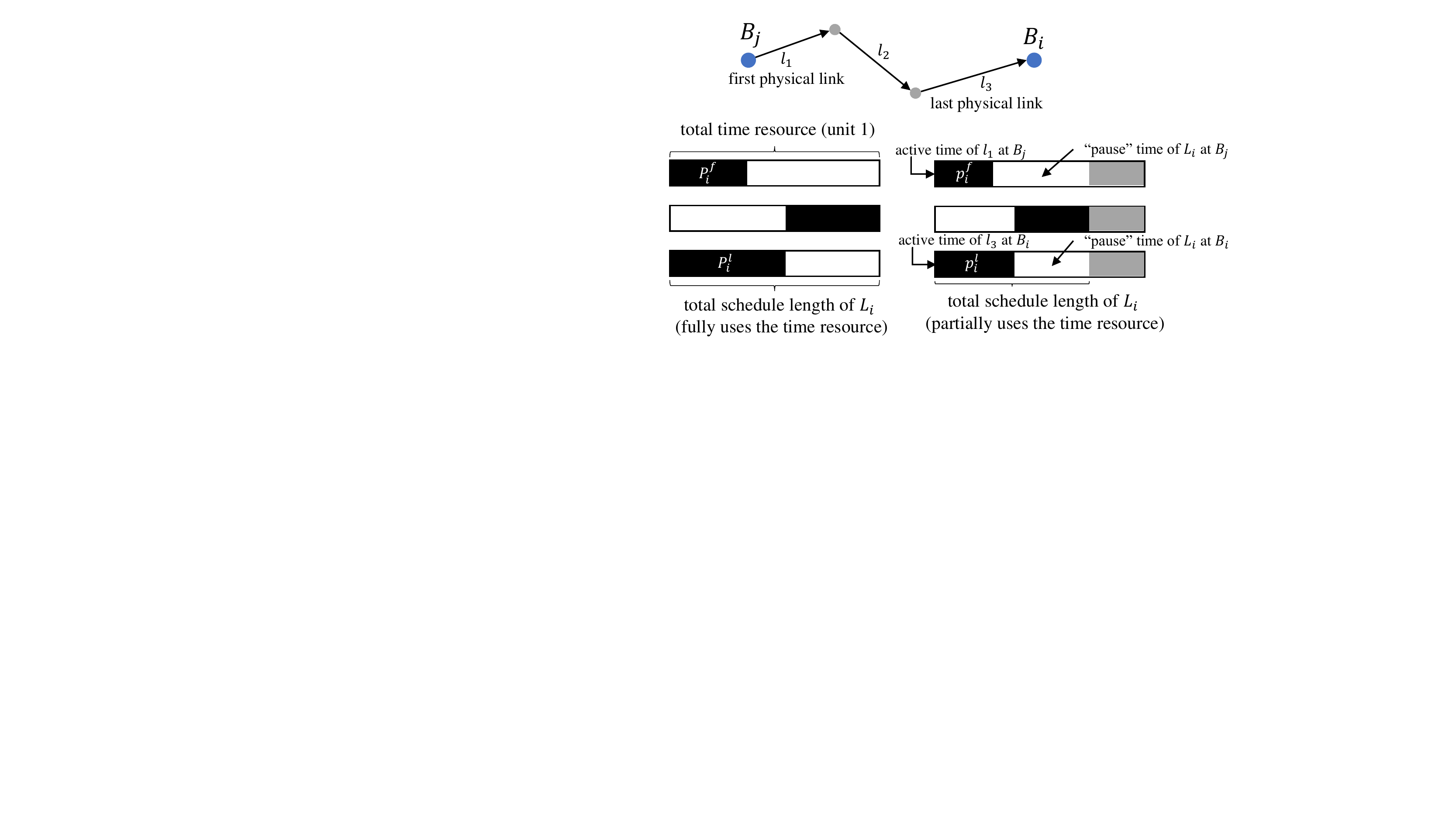}
  \caption{An example of the optimal schedule of a logical link $L_i$}
  \label{fig:example_logical_link_schedule}
\end{figure}

Figure~\ref{fig:example_logical_link_schedule} gives an example of the optimal
schedule of a logical $L_i$. We denote $p_i^f$ and $p_i^l$ as the portion of 
the total time resource when the 
first and last physical links of $L_i$ are scheduled to be active, respectively. 
Therefore, the average logical link rate $c_i$ of $L_i$ over the entire 
time resource can be calculated as,
\begin{equation}
    c_i = \min\{\frac{p_i^f}{P_i^f}, \frac{p_i^l}{P_i^l}\}\cdot C_i
    \label{eq:avg_link_cap}
\end{equation}

Thus, the optimization problem can be written as,
\begin{equation}
\begin{split}
    \max \quad & D_B \\ 
    \mathrm{s.t.} \quad & \min\{\frac{p_i^f}{P_i^f}, \frac{p_i^l}{P_i^l}\}\cdot C_i \geq \sum_{B_j \in \mathcal{B}_{i}}{D_j} = |\mathcal{B}_{i}| D_B,\ \forall\ i \in \{1,2, ..., N_B\} \\
    & 0 \leq p_i^f \leq P_i^f,\ \forall\ i \in \{1,2, ..., N_B\} \\
    & 0 \leq p_i^l \leq P_i^l,,\ \forall\ i \in \{1,2, ..., N_B\} \\
    & p_i^l + \sum_{L_j\in \mathcal{L}_i\backslash L_i}{p_j^f} \leq N_i^R,,\ \forall\ i \in \{1,2, ..., N_B\} \\
    & \sum_{L_i\in \mathcal{L}_M}{p_i^f} \leq N_M^R
\end{split}
\label{eq:optimal_case2}
\end{equation}
where $\mathcal{L}_i$ is the set of logical links attached to the small-cell BS 
$B_i$; $\mathcal{L}_M$ is the set of logical links attached to the macro-cell BS $M$.

As shown in Equation~\ref{eq:optimal_case2}, the logical link capacity constraint is 
different from that in the case 1, because when logical links compete for the radio 
chains, it is possible that a logical link $L_i$ cannot achieve its optimal capacity 
$C_i$ due to less time would be scheduled for the transmission on its first or last 
physical link. In fact, since the schedule of the physical links within a logical 
link follows the logical link's optimal schedule, we can get
\begin{equation*}
    \frac{p_i^f}{P_i^f} = \frac{p_i^l}{P_i^l}
\end{equation*}
because when the bottleneck of a logical link's capacity is determined by either 
its first physical link or its last physical link in Equation~\ref{eq:avg_link_cap}, 
allocating more active time for the physical link at the other end is meaningless.

Therefore, we can use $p_i^f$ to represent $p_i^l$, and update the formation of the 
optimization problem as,
\begin{equation}
\begin{split}
    \max \quad & D_B \\ 
    \mathrm{s.t.} \quad & \frac{p_i^f}{P_i^f}\cdot C_i \geq \sum_{B_j \in \mathcal{B}_{i}}{D_j} = |\mathcal{B}_{i}| D_B,\ \forall\ i \in \{1,2, ..., N_B\} \\
    & 0 \leq p_i^f \leq P_i^f,\ \forall\ i \in \{1,2, ..., N_B\} \\
    & \frac{P_i^l}{P_i^f}\cdot p_i^f + \sum_{L_j\in \mathcal{L}_i\backslash L_i}{p_j^f} \leq N_i^R,\ \forall\ i \in \{1,2, ..., N_B\} \\
    & \sum_{L_i\in \mathcal{L}_M}{p_i^f} \leq N_M^R
\end{split}
\label{eq:optimal_case2_new}
\end{equation}

\subsubsection{Case 3: limited-interference, enough-radio-chain}
\label{sec:lp_case3}

Before moving to the problem formulation for this case, we have make it clear 
that the mutual interference between logical links is produced by the mutual 
interference between physical links within them. To eliminate the existing mutual 
interference between logical links, the straight-forward idea is to schedule the 
interfering physical links into different time periods. However, a logical link 
may contain several physical links, and each physical link may have its individual 
interference relationship with different physical links of other logical links, 
which makes the physical link level interference-avoid scheduling complicated.
Not to mention that it likely breaks the optimal scheduling assumption within each
logical link. 
Therefore, when we deal with the mutual interference between logical links, we 
consider a multi-hop logical link as a ``virtual link", and the schedule of the 
physical links within a logical link follows the optimal schedule of that logical 
link, which is the same assumption as used in case 2. Thus, we only need to make 
sure that the scheduled active time periods of any two interfering logical links 
do not overlap. This is referred to as the \textit{logical link interference constraint}.
Since only the logical link level interference-avoid scheduling is in use to
maximize the traffic demand of small-cell BSs, 
the schedule found by solving the following optimization problem is sub-optimal
in the original problem setting.

The maximize the traffic demand in this case, we have to solve the following linear
programming problem.
\begin{equation}
\begin{split}
    \max \quad & D_B \\ 
    \mathrm{s.t.} \quad & \frac{p_i^f}{P_i^f}\cdot C_i \geq \sum_{B_j \in \mathcal{B}_{i}}{D_j} = |\mathcal{B}_{i}| D_B,\ \forall\ i \in \{1,2, ..., N_B\} \\
    & 0 \leq p_i^f \leq P_i^f,\ \forall\ i \in \{1,2, ..., N_B\} \\
    & \frac{p_i^f}{P_i^f} + I_{ij}\cdot{\frac{p_j^f}{P_j^f}} \leq 1,\ \forall\ i,j \in \{1,2, ..., N_B\} \\
\end{split}
\label{eq:optimal_case3}
\end{equation}
where $I_{ij}$ is a binary parameter identifying the interference relationship between logical 
link $L_i$ and $L_j$. If $L_i$ and $L_j$ interfere with each other, $I_{ij}=1$; otherwise, $I_{ij}=0$. 
From the analysis on the interference relationship between physical links in~\cite{hu2020feasibility}, we can derive that $L_i$ interferes $L_j$ if and only if $L_j$ interferes 
$L_i$, which means $I_{ij}=I_{ji}$. 

Comparing Equation~\ref{eq:optimal_case2_new} and Equation~\ref{eq:optimal_case3}, there is 
similarity between  expressions of the ``radio chain constraint" and the ``logical link 
interference constraint". It is intuitive to think that both of them reflect the idea of 
scheduling the link transmissions to avoid interference in the network, since the ``radio 
chain constraint" can be regarded as the ``primary interference constraint".

\subsubsection{Case 4: limited-interference, limited-radio-chain}
This case can be considered as the combination of case 2 and case 3, as both primary and secondary 
mutual interference have impact on the schedule of transmissions in the backhaul network. 

We can formulate the optimization problem as,
\begin{equation}
\begin{split}
    \max \quad & D_B \\ 
    \mathrm{s.t.} \quad & \frac{p_i^f}{P_i^f}\cdot C_i \geq \sum_{B_j \in \mathcal{B}_{i}}{D_j} = |\mathcal{B}_{i}| D_B,\ \forall\ i \in \{1,2, ..., N_B\} \\
    & 0 \leq p_i^f \leq P_i^f,\ \forall\ i \in \{1,2, ..., N_B\} \\
    & \frac{p_i^f}{P_i^f} + I_{ij}\cdot{\frac{p_j^f}{P_j^f}} \leq 1,\ \forall\ i,j \in \{1,2, ..., N_B\} \\
    & \frac{P_i^l}{P_i^f}\cdot p_i^f + \sum_{L_j\in \mathcal{L}_i\backslash L_i}{p_j^f} \leq N_i^R,\ \forall\ i \in \{1,2, ..., N_B\} \\
    & \sum_{L_i\in \mathcal{L}_M}{p_i^f} \leq N_M^R
\end{split}
\label{eq:max_td_sbs_case4}
\end{equation}
where all the variables and parameters are the same as defined in previous cases.

\subsection{Maximizing the total backhaul traffic demand at the macro-cell BS}
\label{sec:max_td_mbs}
Based on the above analysis, we can easily adapt the problem formation in the 
section~\ref{sec:max_td_sbs} to obtain the optimization problem of maximizing the 
total backhaul traffic demand at the macro-cell BS. In this problem, the traffic
demand of each small-cell BS could be different. We first formulate the problem 
corresponding to the case 1 in the section~\ref{sec:max_td_sbs}.

\subsubsection{Case 1: interference-minimal, enough-radio-chain}

When neither ``intra-path" nor ``inter-path" mutual interference exists and 
there are enough radio chains to support the concurrent transmissions of 
multiple logical links at each BS, the maximum backhaul traffic demand at the 
macro-cell BS can be obtained through solving the following linear programming
problem,
\begin{equation}
    \begin{split}
        \max \quad & D_M = \sum_{i=1}^{N_B}{D_i} \\
        \mathrm{s.t.} \quad & C_i \geq \sum_{B_j \in \mathcal{B}_{i}}{D_j}, \ \forall\ i \in \{1,2, ..., N_B\} \\
    \end{split}
\end{equation}
where all the variables and parameters are the same as defined in the 
section~\ref{sec:max_td_sbs}. 

As the case 4 is the combination of case 2 and 3 in the 
section~\ref{sec:max_td_sbs}, we will only provide the problem formation of 
the case where limited mutual interference exists and each BS has limited 
radio chain resources.

\subsubsection{Case 2: limited-interference, limited-radio-chain}
The optimization problem of this case can be formulated as follow,
\begin{equation}
\begin{split}
    \max \quad & D_M = \sum_{i=1}^{N_B}{D_i} \\ 
    \mathrm{s.t.} \quad & \frac{p_i^f}{P_i^f}\cdot C_i \geq \sum_{B_j \in \mathcal{B}_{i}}{D_j},\ \forall\ i \in \{1,2, ..., N_B\} \\
    & 0 \leq p_i^f \leq P_i^f,\ \forall\ i \in \{1,2, ..., N_B\} \\
    & \frac{p_i^f}{P_i^f} + I_{ij}\cdot{\frac{p_j^f}{P_j^f}} \leq 1,\ \forall\ i,j \in \{1,2, ..., N_B\} \\
    & \frac{P_i^l}{P_i^f}\cdot p_i^f + \sum_{L_j\in \mathcal{L}_i\backslash L_i}{p_j^f} \leq N_i^R,\ \forall\ i \in \{1,2, ..., N_B\} \\
    & \sum_{L_i\in \mathcal{L}_M}{p_i^f} \leq N_M^R
\end{split}
\label{eq:optimal_case2_mbs}
\end{equation}
where all the variables and parameters are the same as defined in the section~\ref{sec:max_td_sbs}.

In the simulation section of this chapter, we focus on the performance evaluation of the cases where the traffic demand on each small-cell BS is the same; while leave the performance evaluation of the cases discussed in this section as our future work.

\subsection{Maximizing the total backhaul throughput while considering fairness}
\label{sec:max_td_mbs_fair}
As we can see from the numerical results in Figure~\ref{fig:fairness_traffic_demand}, the traffic demand of each 
small-cell BS in the backhaul network is very unbalanced, when the aggregated traffic 
demand achieves the maximum at the macro-cell BS. In the cellular system, the resources 
have to be allocated in a way that fairness is considered to some extent, so that every 
cellular user who is paying their monthly bill can receive a service meeting the minimum 
quality of service requirement. From this perspective, we are interested in maximizing the
total backhaul traffic demand aggregated at the macro-cell BS while considering the fairness issue across all the small-cell BSs in the backhaul network. Specifically, we complete the task through the following two steps.

First, we take advantage of the work in the section~\ref{sec:max_td_sbs}, to maximize the equal traffic demand of all small-cell BSs in the backhaul network. The maximum traffic demand value found using the linear programming formulation in Equation~\ref{eq:max_td_sbs_case4} will be served as the minimum traffic demand of each small-cell BS that is able to be supported by the given backhaul network. Then we try to utilize the available network resources to accommodate more traffic demand on some of the small-cell BSs. Here we only use the formulation of case 4 in section~\ref{sec:max_td_sbs}, because the other three cases can be regarded as special cases of case 4. 

Second, we use the similar idea in section~\ref{sec:max_td_mbs} to maximize the total aggregated backhaul demand at the macro-cell BS, while the minimum traffic demand of each small-cell BS obtained in the above first step can be satisfied. The objective can be achieved by solving the following linear programming formulation,

\begin{equation}
\begin{split}
    \max \quad & D_M = \sum_{i=1}^{N_B}{D_i} \\ 
    \mathrm{s.t.} \quad & \frac{p_i^f}{P_i^f}\cdot C_i \geq \sum_{B_j \in \mathcal{B}_{i}}{D_j},\ \forall\ i \in \{1,2, ..., N_B\} \\
    & D_i \geq \widehat{D_B}, \ \forall\ i \in \{1,2, ..., N_B\} \\
    & 0 \leq p_i^f \leq P_i^f,\ \forall\ i \in \{1,2, ..., N_B\} \\
    & \frac{p_i^f}{P_i^f} + I_{ij}\cdot{\frac{p_j^f}{P_j^f}} \leq 1,\ \forall\ i,j \in \{1,2, ..., N_B\} \\
    & \frac{P_i^l}{P_i^f}\cdot p_i^f + \sum_{L_j\in \mathcal{L}_i\backslash L_i}{p_j^f} \leq N_i^R,\ \forall\ i \in \{1,2, ..., N_B\} \\
    & \sum_{L_i\in \mathcal{L}_M}{p_i^f} \leq N_M^R
\end{split}
\label{eq:max_td_mbs_fair}
\end{equation}
where $\widehat{D_B}$ is the maximum equal traffic demand obtained through the first step; while the other parameters and variables are the same as defined in the previous formulations.

\section{Schedule the maximum traffic demand in the relay-assisted mmWave backhaul network}
\label{sec:6_4}
After solving the optimization problems described in the above section, we need to figure out 
how to schedule the transmission of each link in the backhaul network, so that the maximum 
traffic demand $D_B$ at each small-cell BS or the maximum backhaul throughput $D_M$ at the 
macro-cell BS can be achieved.
As the by-product of maximizing traffic demand of BSs using linear programming, we can obtain a
set of $\{p_i^f\}$ values, where each $p_i^f$ represents the portion of the total time resource
expected to be assigned to the first physical link of a logical link $L_i$. 
We propose a `depth-first"-style algorithm to find a schedule accommodating all the values in
$\{p_i^f\}$ for all logical links in the backhaul network.

The algorithm described using the pseudo code in the Algorithm~\ref{alg:6_4} aims to address
the scheduling issue in the case where limited mutual interference exists and there is limited 
radio chain resource on each BS in a tree-style relay-assisted mmWave backhaul network. Since 
other cases are special cases of this one, the algorithm can easily be modified to adapt to them.
Since the optimization problem can deal with either downlink or uplink traffic scenario, the 
scheduling algorithm can handle either downlink or uplink traffic scenario as well, because the
operations in the algorithm are not affected by the change of traffic directions in the network.

Basically, the scheduling process starts from the macro-cell BS $M$, runs in a ``depth-first" way, 
and does not end until all non-leaf small-cell BSs have made their scheduling decisions. Every
BS in the backhaul network only has to make scheduling decisions for all the logical links between 
the BS and its ``child" BSs, because the schedule of the logical link between a BS and its 
``parent" BS has already been decided by the ``parent" BS.

\begin{algorithm}[ht!]
    \caption{Optimal scheduling algorithm for tree-style mmWave backhaul networks}
    \label{alg:6_4}
    \algsetup{linenosize=\large}
    \begin{algorithmic}[1]
        \REQUIRE $\{p_i^f\}$
        \STATE $\mathcal{S}.\mathrm{push}(M)$;
        \WHILE{$\mathcal{S}\neq \emptyset$}
            \STATE $B_c = \mathcal{S}.\mathrm{pop}()$;
            \STATE $\mathcal{L}_c\gets \emptyset$;
            \FOR{each $B_i\in \mathcal{B}\land (B_i\ \mathrm{is\ a\ child\ of}\ B_c)$}
                \STATE $\mathcal{S}.\mathrm{push}(B_i)$;
                \STATE $\mathcal{L}_c.\mathrm{insert}(L_i)$;
            \ENDFOR
            \IF{$B_c\neq M$}
                \STATE assign the schedule length of $\frac{p_c^f}{P_c^f}$ on the $1^{st}$ radio chain according to the existing schedule of $L_c$;
                \IF{$L_c$ interferes with a single $L_i\in \mathcal{L}_c$}
                    \STATE assign the schedule length of $\frac{p_i^f}{P_i^f}$ on the $1^{st}$ radio chain for $L_i$, which does not overlap with the schedule of $L_c$;
                    \STATE $\mathcal{L}_c.\mathrm{remove}(L_i)$;
                \ENDIF
            \ENDIF
            \WHILE{there exists $L_i\in \mathcal{L}_c\ \land\ L_i$ does not interfere with any other logical link in $\mathcal{L}_c$}
                \STATE assign the schedule length of $p_i^f$ on the first available radio chain where blank time periods exist for $L_i$.
                \IF{the first available radio chain does not have enough time available for $L_i$}
                    \STATE assign the rest of the schedule length of $L_i$ in the next available radio chain, and the assigned time period should not overlap with the already scheduled time period of $L_i$ on the previous radio chain.
                \ENDIF
                \STATE $\mathcal{L}_c.\mathrm{remove}(L_i)$;
            \ENDWHILE
            \WHILE{there exist $L_i,\ L_j\in\ \mathcal{L}_c\ \land\ L_i$ interferes with $L_j$}
                \STATE first assign the time periods $p_i^f$ and $p_j^f$ for the first physical links of $L_i$ and $L_j$, respectively, on the first available radio chain where blank time periods exist.
                \IF{the first available radio chain does not have enough time available}
                    \STATE assign the rest of the schedule length of the first physical links of $L_i$ and $L_j$ in the next available radio chain, and the assigned time period should not overlap with the already scheduled time periods of them on the previous radio chain.
                \ENDIF
                \STATE assign the ``pause" time periods $p_i^f\cdot (\frac{1-P_i^f}{P_i^f})$ and $p_j^f\cdot (\frac{1-P_j^f}{P_j^f})$ on the radio chains where the first physical links of $L_i$ and $L_j$ have been scheduled. These ``pause" time periods should not overlap with each other or the scheduled periods of their corresponding first physical links, but they must reuse the time periods scheduled for other logical links, if there exist.
                \STATE $\mathcal{L}_c.\mathrm{remove}(L_i, L_j)$;
            \ENDWHILE
        \ENDWHILE
    \end{algorithmic}
\end{algorithm}

As shown in Algorithm~\ref{alg:6_4}, a stack $\mathcal{S}$ stores the set of BSs 
whose scheduling decisions need to be made. When one BS $B_c$ is in turn to make its 
scheduling decision, a set $\mathcal{L}_c$ is used to store the to be scheduled logical
links. If $B_c$ is a small-cell BS, the schedule of the logical link $L_c$ between $B_c$'s 
parent BS and $B_c$ has already been determined by $B_c$'s parent BS. Therefore, $B_c$ will 
first allocate time for $L_c$ according to the determined schedule. If there exists one 
logical link to be scheduled interfering with $L_c$, $B_c$ schedules the interfering logical 
link right after scheduling $L_c$. After that, $B_c$ will schedule all non-interfering 
logical links, whose scheduled time periods do not overlap with each other, but can occupy
the ``pause" time of the previously scheduled logical links. Note that the schedule length of 
a logical link at a BS contains two parts: one is the scheduled time period for the physical 
link attached to the BS, and the other is the ``pause" time which is used by other physical 
links within the logical link (see Figure~\ref{fig:example_logical_link_schedule}). However, the logical link does not need to use the actual radio 
chain resource at the BS during the ``pause" time, as the physical links being active at the 
``pause" time are not attached to that BS. In the end, $B_c$ schedule the remaining logical 
links interfering with each other in a pair-wise way. For a pair of interfering logical links,
$B_c$ first assign the ``active" time for their physical links attached to $B_c$, and then 
assign the ``pause" time. Note that ``pause" time of a logical link $L_i$ can overlap with the 
schedule of any logical link that is not interfering with $L_i$.


\section{Numerical results and analysis}
This section aims to provide several sets of numerical results and simulation results, which will be analyzed to show the throughput performance of the relay-assisted mmWave backhaul network.

\subsection{The maximum traffic demand of each small-cell BS}

\begin{figure}[ht]
  \centering
  \includegraphics[width=0.9\linewidth]{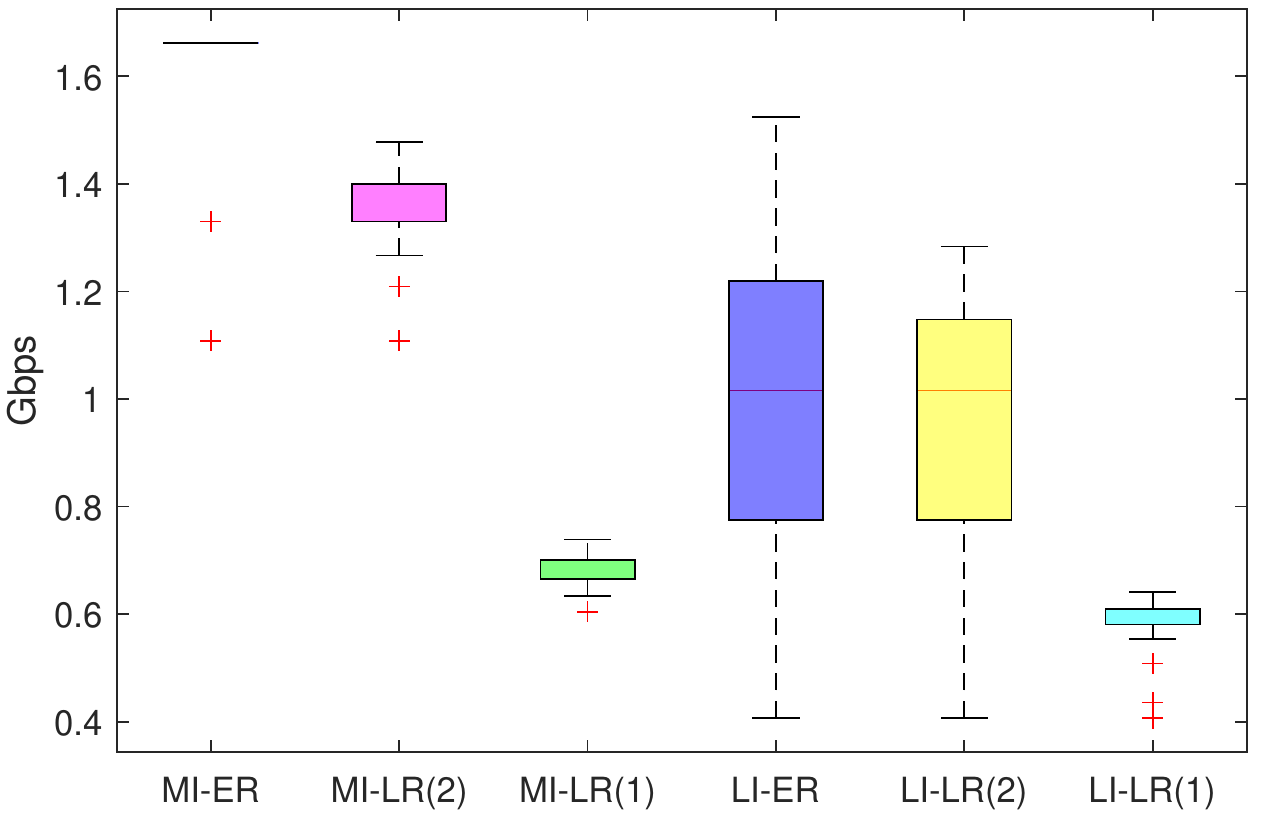}
  \caption{Maximum traffic demand of each small-cell BS in different network settings}
  \label{fig:max_demand}
\end{figure}

First, the maximum traffic demand of each small-cell BS in a mmWave backhaul network is calculated in different network settings using the method introduced in section~\ref{sec:ch6:max_traffic}. Note that, in the calculation, only downlink traffic or uplink traffic is considered. 
Nowadays, the communication systems usually apply the OFDM and QAM modulation schemes. In our
simulations, we consider a frame structure with the slot size 4.16 $\mu s$, a total bandwidth 2 GHz, which contains 6912 sub-carriers, and the 256 QAM is in use. Therefore, the raw data rate \footnote{Here, we do not consider the channel coding efficiency.} of each physical link is fixed at $6912\times 8 / 4.16 = 13.3\ \mathrm{Gbps}$. Therefore, the data rate of a single-hop logical link equals to the physical link data rate, i.e., 13.3 Gbps; while the data rate of a multi-hop logical link equals to half of the physical link data rate, i.e., 6.65 Gbps. As shown in Figure~\ref{fig:max_demand}, six different network settings are considered, in which ``MI-ER" refers to the case where interference is minimal and enough radio chains are available, ``MI-LR" refers to interference-minimal and limited radio chains, ``LI-ER" refers to the case where interference exists between a few pairs of logical links and enough radio chains are available, while ``LI-LR" refers to the case of limited interference and limited radio chains. In the cases of enough radio chains, the number of radio chains of each BS is set to the number of logical links attached to it; while in the cases of limited radio chains, each small-cell BS has only 1 radio chain, but the macro-cell BS has 2 or 1 radio chains identified by the number in the parenthesis at the end of the label. From Figure~\ref{fig:max_demand}, we can see that the number of radio chains plays a crucial role in determine the maximum traffic demand of each small-cell BS, as the average values of the cases where enough radio chains are available are much higher than that of the cases where limited radio chains are available. One main reason is that in the considered topology, many logical links (e.g., 8) are attached to the macro-cell BS, and if only 1 radio chain is available, at any time, there is at most 1 logical link actively transmitting data, which is very inefficient. We also observe that when the number of radio chains at the macro-cell increases to 2, the obtained maximum traffic demand also increases a lot.  Among the cases with limited radio chains, when interference exists between a few pairs of logical links, the maximum traffic demand drops slightly, because the primary interference only restricts the physical links attached to a BS of different logical links cannot be active simultaneously, while the mutual interference applies a even tighter constraint that the schedules of interfering logical links do not overlap with each other, and the schedule of a multi-hop logical link is much longer than that of a physical link. Similarly, the maximum traffic demand in the ``MI-ER" case is much better than that in the ``LI-ER" case, as the mutual interference is minimal in the ``MI-ER" case. In Figure~\ref{fig:max_demand}, we can observe that the dynamic range of the values in the ``LI-ER" case is very large. It is because that the maximum traffic demand is greatly affected by where the interference appears. If the most of the interfering pairs of logical links are attached to the macro-cell BS, the throughput will be very low; however, if the interference appears mostly at small-cell BSs, as the traffic demand around 2 Gbps can be easily handled when there is a few logical links (e.g., 2) attached to the small-cell BS, which is exactly the case in the considered topologies. 

\subsection{The lower bound of ``enough" radio chains on each BS}
As soon as the maximum traffic demand of each small-cell BS is calculated in the cases where enough radio chains are available on each BS, the portion $p_i^f$ of the schedule of the first physical link within the logical link $L_i$ attached to the parent BS of $B_i$ can also be determined. Based on this set of values, the minimum number of radio chains needed at each BS can be updated. 

\begin{table}[h!]
	\caption{The numerical result on the minimum number of ``enough" radio chains}
	\centering
	\begin{tabular}{|c|c|c|} 
		\hline 
		BS type & interference-minimal & limited-interference \\ 
		\hline 
		small-cell & 1 (60$\%$), 2 (40\%) & 1 (62\%), 2 (38\%)\\
		\hline
		macro-cell &  2 (32$\%$), 3 (68$\%$) & 1 (5\%), 2 (69\%), 3 (26\%)\\ 
		\hline 
	\end{tabular}
	\label{tb:enough_rr}
\end{table}

As shown in Table~\ref{tb:enough_rr}, in every tested case, only single radio chain is enough for the main portion of all small-cell BSs, no matter there is mutual interference or not. However, for the small-cell BSs close to the macro-cell BS, they may need two radio chains. This is because in most of cases, each small-cell BS may only have at most 2 logical links attached, and the BSs close to the macro-cell BS tend to have larger aggregated traffic demand. However, as for the macro-cell BS, since usually 8 logical links are attached to it, more radio chains are needed. But 2-3 radio chains are enough for the macro-cell BS in all test cases, which is much smaller than the initial ``enough" number of radio chains determined by the number of logical links attached (i.e., 8). This result is reasonable, because most of the logical links are multi-hop, and in their optimal schedule, each physical link can intuitively be thought as active for only half of the total schedule of the logical link. Thus, single radio chain can serve 2 logical links without any problem if they do not interfere with each other. 

\subsection{The maximum aggregated traffic demand at the macro-cell BS}
Simulations are also conducted to show the aggregated traffic demand values at the macro-cell BS upon three different maximization objectives in two different network setting configurations. 
\begin{figure}[ht]
  \centering
  \includegraphics[width=0.9\linewidth]{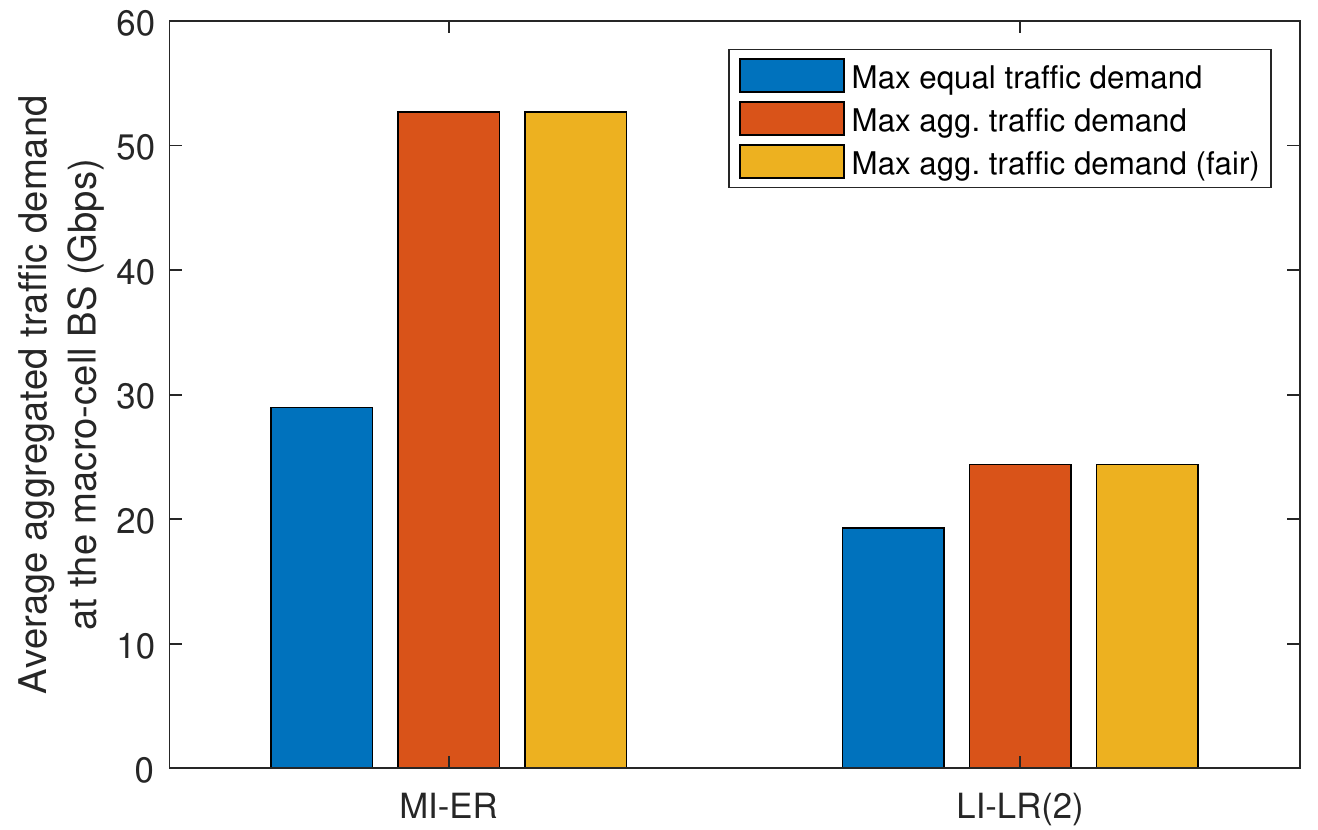}
  \caption{Maximum aggregated traffic demand at the macro-cell BS}
  \label{fig:max_agg_traffic_demand}
\end{figure}
In Figure~\ref{fig:max_agg_traffic_demand}, the blue column represents the value corresponding to the case where equal traffic demand of every small-cell BS is maximized; the red column represents the value corresponding to the case where the aggregated traffic demand at the macro-cell BS is maximized without considering fairness; while the yellow column represents the case where fairness is considered when the aggregated traffic demand is maximized. The left three columns corresponds to the values in the interference-minimal backhaul networks with enough radio chain resources; while the right three columns shows the values in the backhaul networks where limited interference exists between a few pairs of logical links, and limited radios are available on BSs (i.e., 2 radio chains on the macro-cell BS, and 1 radio chain on each small-cell BS). Each value shown in the figure is an averaged value across 50 sets of individual simulations using the backhaul network data generated in~\cite{hu2020feasibility}.

As shown in Figure~\ref{fig:max_agg_traffic_demand}, maximizing the aggregated traffic demand at the macro-cell BS allows more traffic to flow into the mmWave backhual network than maximizing the equal traffic demand of each small-cell BS does, because in the later case, the equal traffic demand is limited by the traffic demand of BSs on the ``bottleneck" routes. The bottleneck routes are the routes consisting the largest number of small-cell BSs in the backhual network. We can also see that the gap of aggregated traffic demand between the blue column and the yellow column shrinks from the MI-ER case to the LI-LR(2) case. This is because in the MI-ER case, plenty of radio chains are available on each BS, especially on the macro-cell BS, which means more extra resources are available after allocating the time resources to small-cell BSs to achieve the maximum equal traffic demand of each small-cell BS, and with more available extra resources, more additional traffic can flow into the backhaul network. It is also interesting to see that from the simulation result, considering the fairness factor in maximizing the aggregated traffic demand does not affect the achievable maximum value of aggregated traffic demand at the macro-cell BS, because in both MI-ER and LI-LR(2) cases, the logical links attached to the macro-cell BS can achieve the same utilization of the available radio chains on the macro-cell BS.     
\begin{figure}[h!]
  \centering
  \includegraphics[width=0.9\linewidth]{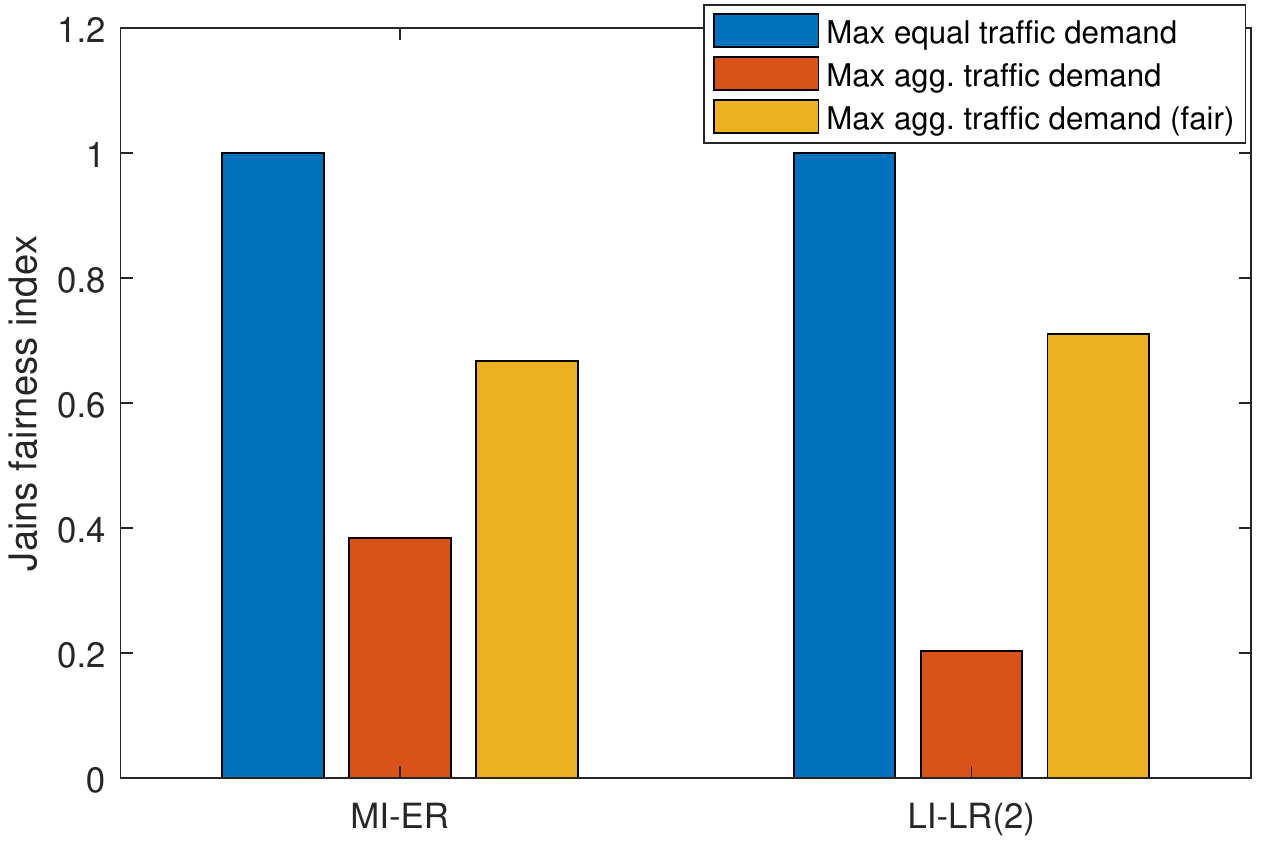}
  \caption{Jain's fairness index on the traffic demand of small-cell BSs}
  \label{fig:fairness_traffic_demand}
\end{figure}

On the other hand, from the perspective of fairness, Figure~\ref{fig:fairness_traffic_demand} shows the comparison among different scenarios in terms of the Jain's fairness index~\cite{jain1999throughput}. It is obvious that when the equal traffic demand of each small-cell BS is maximized, the fairness index reaches its highest value of 1. Meanwhile, if fairness is not considered, purely maximizing the aggregated traffic demand at the macro-cell BS leads to poor fairness, which are shown by the red columns in the figure. However, as the fairness factor is considered as introduced in section~\ref{sec:max_td_mbs_fair}, the fairness index value increases significantly, which is shown by the yellow columns.

\section{Conclusion}
In this paper, the throughput performance issue of the relay-assisted mmWave backhaul network is investigated. The maximum traffic demand of small-cell BSs and the maximum throughout at the macro-cell BS in the tree-style backhaul network have been found using linear programming under different network settings, which concern both the number of radio chains available on BSs and the interference relationship between logical links in the backhaul network. A new interference model for the relay-assisted mmWave backhaul network in the dense urban environment is proposed, which reflects the limited interference footprint of mmWave directional communications. 

\bibliographystyle{ieeetr}
\bibliography{sample}

\begin{thebibliography}{10}

\bibitem{pi2011introduction}
Z.~Pi and F.~Khan, ``An introduction to millimeter-wave mobile broadband
  systems,'' {\em IEEE communications magazine}, vol.~49, no.~6, 2011.

\bibitem{weiler2014enabling}
R.~J. Weiler, M.~Peter, W.~Keusgen, E.~Calvanese-Strinati, A.~De~Domenico,
  I.~Filippini, A.~Capone, I.~Siaud, A.-M. Ulmer-Moll, A.~Maltsev, {\em
  et~al.}, ``Enabling {5G} backhaul and access with millimeter-waves,'' in {\em
  Networks and Communications (EuCNC), 2014 European Conference on}, pp.~1--5,
  IEEE, 2014.

\bibitem{verma2015backhaul}
L.~Verma, M.~Fakharzadeh, and S.~Choi, ``Backhaul need for speed: 60 {GHz} is
  the solution,'' {\em IEEE Wireless Communications}, vol.~22, no.~6,
  pp.~114--121, 2015.

\bibitem{jaber20165g}
M.~Jaber, M.~A. Imran, R.~Tafazolli, and A.~Tukmanov, ``{5G} backhaul
  challenges and emerging research directions: {A} survey,'' {\em IEEE Access},
  vol.~4, pp.~1743--1766, 2016.

\bibitem{yan2021load}
Y.~Yan, Q.~Hu, and D.~M. Blough, ``Load-balanced routing for hybrid
  fiber/wireless backhaul networks,'' in {\em 2021 IEEE Global Communications
  Conference (GLOBECOM)}, pp.~1--6, IEEE, 2021.

\bibitem{frey1999effects}
T.~L. Frey, ``The effects of the atmosphere and weather on the performance of a
  {mm-Wave} communication link,'' {\em Applied Microwave and Wireless},
  vol.~11, pp.~76--81, 1999.

\bibitem{liu2019analysis}
Y.~Liu and D.~M. Blough, ``Analysis of blockage effects on roadside
  relay-assisted mmwave backhaul networks,'' in {\em ICC 2019-2019 IEEE
  International Conference on Communications (ICC)}, pp.~1--7, IEEE, 2019.

\bibitem{singh2009blockage}
S.~Singh, F.~Ziliotto, U.~Madhow, E.~Belding, and M.~Rodwell, ``Blockage and
  directivity in 60 {GHz} wireless personal area networks: {From} cross-layer
  model to multihop {MAC} design,'' {\em IEEE Journal on Selected Areas in
  Communications}, vol.~27, no.~8, 2009.

\bibitem{hu2017relay}
Q.~Hu and D.~M. Blough, ``Relay selection and scheduling for millimeter wave
  backhaul in urban environments,'' in {\em 2017 IEEE 14th International
  Conference on Mobile Ad Hoc and Sensor Systems (MASS)}, pp.~206--214, IEEE,
  2017.

\bibitem{hu2018optimizing}
Q.~Hu and D.~M. Blough, ``Optimizing millimeter-wave backhaul networks in
  roadside environments,'' in {\em Communication (ICC), 2018 IEEE International
  Conference on}, IEEE, 2018.

\bibitem{hu2020feasibility}
Q.~Hu and D.~M. Blough, ``On the feasibility of high throughput mmwave backhaul
  networks in urban areas,'' in {\em 2020 International Conference on
  Computing, Networking and Communications (ICNC)}, pp.~572--578, IEEE, 2020.

\bibitem{yan2018path}
Y.~Yan, Q.~Hu, and D.~M. Blough, ``Path selection with amplify and forward
  relays in {mmWave} backhaul networks,'' in {\em 2018 IEEE 29th Annual
  International Symposium on Personal, Indoor and Mobile Radio Communications
  (PIMRC)}, pp.~1--6, IEEE, 2018.

\bibitem{yan2021feasibility}
Y.~Yan, Q.~Hu, and D.~M. Blough, ``Feasibility of multipath construction in
  mmwave backhaul,'' in {\em 2021 IEEE 22nd International Symposium on a World
  of Wireless, Mobile and Multimedia Networks (WoWMoM)}, pp.~81--90, IEEE,
  2021.

\bibitem{liu2020joint}
Y.~Liu, Q.~Hu, and D.~M. Blough, ``Joint link-level and network-level
  reconfiguration for urban mmwave wireless backhaul networks,'' {\em Computer
  Communications}, vol.~164, pp.~215--228, 2020.

\bibitem{yang2009relay}
Y.~Yang, H.~Hu, J.~Xu, and G.~Mao, ``Relay technologies for {WiMAX} and
  {LTE}-advanced mobile systems,'' {\em IEEE Communications Magazine}, vol.~47,
  no.~10, pp.~100--105, 2009.

\bibitem{peters2009relay}
S.~W. Peters, A.~Y. Panah, K.~T. Truong, and R.~W. Heath, ``Relay architectures
  for {3GPP} {LTE}-advanced,'' {\em EURASIP Journal on Wireless Communications
  and Networking}, vol.~2009, no.~1, p.~618787, 2009.

\bibitem{lan2011space}
Z.~Lan, L.~A. Lu, X.~Zhang, C.~Pyo, and H.~Harada, ``A space-time scheduling
  assisted cooperative relay for {mmWave WLAN/WPAN} systems with directional
  antenna,'' in {\em Global Telecommunications Conference (GLOBECOM 2011), 2011
  IEEE}, pp.~1--6, IEEE, 2011.

\bibitem{zheng2016toward}
G.~Zheng, C.~Hua, R.~Zheng, and Q.~Wang, ``Toward robust relay placement in 60
  {GHz} mmwave wireless personal area networks with directional antenna,'' {\em
  IEEE Transactions on Mobile Computing}, vol.~15, no.~3, pp.~762--773, 2016.

\bibitem{niu2015blockage}
Y.~Niu, Y.~Li, D.~Jin, L.~Su, and D.~Wu, ``Blockage robust and efficient
  scheduling for directional {mmWave WPANs},'' {\em IEEE Transactions on
  Vehicular Technology}, vol.~64, no.~2, pp.~728--742, 2015.

\bibitem{qiao2012efficient}
J.~Qiao, B.~Cao, X.~Zhang, X.~Shen, and J.~W. Mark, ``Efficient concurrent
  transmission scheduling for cooperative millimeter wave systems,'' in {\em
  Global Communications Conference (GLOBECOM), 2012 IEEE}, pp.~4187--4192,
  IEEE, 2012.

\bibitem{arribas2019optimizing}
E.~Arribas, A.~F. Anta, D.~R. Kowalski, V.~Mancuso, M.~A. Mosteiro, J.~Widmer,
  and P.~W. Wong, ``Optimizing mmwave wireless backhaul scheduling,'' {\em IEEE
  Transactions on Mobile Computing}, vol.~19, no.~10, pp.~2409--2428, 2019.

\bibitem{saad2019millimeter}
M.~Saad and S.~Abdallah, ``On millimeter wave 5g backhaul link scheduling,''
  {\em IEEE Access}, vol.~7, pp.~76448--76457, 2019.

\bibitem{niu2019relay}
Y.~Niu, W.~Ding, H.~Wu, Y.~Li, X.~Chen, B.~Ai, and Z.~Zhong, ``Relay-assisted
  and qos aware scheduling to overcome blockage in mmwave backhaul networks,''
  {\em IEEE Transactions on Vehicular Technology}, vol.~68, no.~2,
  pp.~1733--1744, 2019.

\bibitem{fang2021joint}
C.~Fang, C.~Madapatha, B.~Makki, and T.~Svensson, ``Joint scheduling and
  throughput maximization in self-backhauled millimeter wave cellular
  networks,'' in {\em 2021 17th International Symposium on Wireless
  Communication Systems (ISWCS)}, pp.~1--6, IEEE, 2021.

\bibitem{ranjantwo}
S.~Ranjan, P.~Jha, A.~Karandikar, and P.~Chaporkar, ``Two stage downlink
  scheduling for balancing qos in multihop iab networks,''

\bibitem{yuan2018optimal}
D.~Yuan, H.-Y. Lin, J.~Widmer, and M.~Hollick, ``Optimal joint routing and
  scheduling in millimeter-wave cellular networks,'' in {\em IEEE INFOCOM
  2018-IEEE Conference on Computer Communications}, pp.~1205--1213, IEEE, 2018.

\bibitem{dogan2021reinforcement}
M.~G. Dogan, Y.~H. Ezzeldin, C.~Fragouli, and A.~W. Bohannon, ``A reinforcement
  learning approach for scheduling in mmwave networks,'' in {\em MILCOM
  2021-2021 IEEE Military Communications Conference (MILCOM)}, pp.~771--776,
  IEEE, 2021.

\bibitem{zhang2021resource}
B.~Zhang, F.~Devoti, I.~Filippini, and D.~De~Donno, ``Resource allocation in
  mmwave 5g iab networks: A reinforcement learning approach based on column
  generation,'' {\em Computer Networks}, vol.~196, p.~108248, 2021.

\bibitem{yonis2012lte}
A.~Yonis, M.~Abdullah, and M.~Ghanim, ``{LTE-FDD} and {LTE-TDD} for cellular
  communications,'' {\em Proceeding, Progress in}, 2012.

\bibitem{ku2011resource}
G.~Ku, ``Resource allocation in {LTE},'' {\em Adaptive Signal Processing and
  Information Theory Research Group}, 2011.

\bibitem{scheme2009lte}
B.~T. Scheme, ``{LTE}: the evolution of mobile broadband,'' {\em IEEE
  Communications magazine}, vol.~45, pp.~44--51, 2009.

\bibitem{coleman2018cwna}
D.~D. Coleman and D.~A. Westcott, {\em {CWNA} Certified Wireless Network
  Administrator Study Guide: Exam {CWNA}-107}.
\newblock John Wiley \& Sons, 2018.

\bibitem{ramanathan1997unified}
S.~Ramanathan, ``A unified framework and algorithm for {(T/F/C) DMA} channel
  assignment in wireless networks,'' in {\em Proceedings of INFOCOM'97},
  vol.~2, pp.~900--907, IEEE, 1997.

\bibitem{ergen2010tdma}
S.~C. Ergen and P.~Varaiya, ``{TDMA} scheduling algorithms for wireless sensor
  networks,'' {\em Wireless Networks}, vol.~16, no.~4, pp.~985--997, 2010.

\bibitem{yi2008complexity}
Y.~Yi, A.~Prouti{\`e}re, and M.~Chiang, ``Complexity in wireless scheduling:
  {Impact} and tradeoffs,'' in {\em Proceedings of the 9th ACM international
  symposium on Mobile ad hoc networking and computing}, pp.~33--42, ACM, 2008.

\bibitem{sharma2006complexity}
G.~Sharma, R.~R. Mazumdar, and N.~B. Shroff, ``On the complexity of scheduling
  in wireless networks,'' in {\em Proceedings of the 12th annual international
  conference on Mobile computing and networking}, pp.~227--238, ACM, 2006.

\bibitem{moscibroda2007optimal}
T.~Moscibroda, R.~Rejaie, and R.~Wattenhofer, ``How optimal are wireless
  scheduling protocols?,'' in {\em IEEE INFOCOM 2007-26th IEEE International
  Conference on Computer Communications}, pp.~1433--1441, IEEE, 2007.

\bibitem{blough2008framework}
D.~M. Blough, S.~Das, G.~Resta, and P.~Santi, ``A framework for joint
  scheduling and diversity exploitation under physical interference in wireless
  mesh networks,'' in {\em 2008 5th IEEE International Conference on Mobile Ad
  Hoc and Sensor Systems}, pp.~396--403, IEEE, 2008.

\bibitem{blough2010approximation}
D.~M. Blough, G.~Resta, and P.~Santi, ``Approximation algorithms for wireless
  link scheduling with {SINR}-based interference,'' {\em IEEE/ACM Transactions
  on Networking (ToN)}, vol.~18, no.~6, pp.~1701--1712, 2010.

\bibitem{jain1999throughput}
R.~Jain, A.~Durresi, and G.~Babic, ``Throughput fairness index: An
  explanation,'' in {\em ATM Forum contribution}, vol.~99, 1999.

\end{thebibliography}

\end{document}